\documentclass{elsart}

\usepackage{epsfig}

\usepackage{amssymb}

\begin{document}

\begin{frontmatter}



\title{An MT-Style Optical Package for Optical Data Transmission}

\author{K.K.~Gan}

\address{Department of Physics,
        The Ohio State University,
        Columbus, OH 43210,
        U.S.A.}

\begin{abstract}
An optical package for mounting VCSEL and PIN diodes for transmitting and
receiving optical signals has been developed.
The diodes couple to the fibers in a commercial MT connector.
The package is quite compact with its physical size significantly smaller
than that of the MT connector.
This design simplifies the testing and assembly of the optical components
because the MT connector with the long fibers attached can be remounted
with ease while preserving good light coupling efficiency.
\end{abstract}

\end{frontmatter}

\section{Introduction}
\label{Introduction}
Optical fibers are now replacing traditional copper wires for signal
transmission in the detectors for high energy and nuclear physics research.
Optical fibers are more compact than copper wires, freeing up valuable
detector space, and allow signal transmission over long distances with
small attenuation.
More importantly, the fibers eliminate electromagnetic interference and
ground loops.

Commercially available VCSEL (Vertical Cavity Surface Emitting Laser)
or PIN packages for transmitting and receiving optical signals are quite bulky.
Detectors for high energy and nuclear physics research using optical packages
usually need to custom design their packages due to space constraints.
The problem is especially severe for the vertex detector located
within the vicinity of the beam colliding region. 
In this paper, we present a compact MT-style optical package for housing
VCSEL and PIN diodes.
The diodes couple to the fibers in a commercial MT connector.
This design simplifies the testing and assembly of the optical components
because the MT connector with the long fibers attached can be remounted
with ease while preserving good light coupling efficiency.
In the next section, we present the design of the package.
This is followed by a section on the prototype results and then a summary.

\section{Optical Package Design}
\begin{figure}
\begin{center}
\includegraphics*[width=10cm]{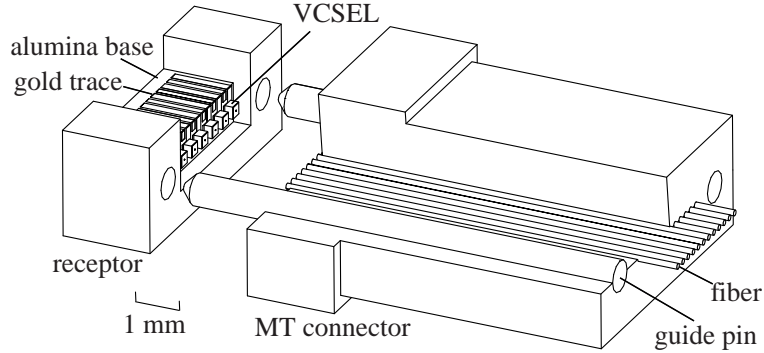}
\end{center}
\caption{
An optical package together with a cut-out view of an MT connector.
The six VCSELs couple to only half of the 12 fibers.}
\label{fig:design}
\end{figure}

The optical package is a two-piece design as shown in Fig.~\ref{fig:design}.
The U-shaped receptor contains two holes for the guide pins.
Each hole has a diameter of $700 \pm 1~\mu$m and the distance between
the two holes is $4600 \pm 3~\mu$m.
The tolerances for the two dimensions are from the MT connector
specification~\cite{tolerances}, hence the same tolerances should be
used in the optical package fabrication.
The receptor can be made of plastic with mold injection.

The base inside the receptor is fabricated using alumina.
The two pieces are glued together with epoxy.
On the side of the base facing the fibers, there are two gold traces
for each VCSEL or PIN diode.
One of the traces is for the cathode and the other for the anode.
A diode is attached to one of the traces with conducting epoxy and
a wire bond connects the other trace to the pad on the top of the diode.
The two traces go over the corner to the other side of the base and
connect via wire bonds to the optical transmitter or receiver chip
placed in close proximity.
The top traces are 1.5~mm in length, hence the chip is connected
to the diode with small stray capacitance and inductance, minimizing
degradation of the electrical signal.
This is one of the advantages of this optical package design.
The base should be placed at a location such that the distance between
the diode and the fiber is $\sim 200~\mu$m for good light
coupling efficiency.

To fabricate the bases, alumina sheet is ground to the proper
thickness of the bases and then cut into strips for deposition
of three-dimensional traces~\cite{HT}.
The trace width is 150~$\mu$m and the gap between traces is 75~$\mu$m.
Most of the deposited traces ($\sim$ 94\%) have good connectivity
across the corner of the strips.
Strips with a large number of traces of good connectivity
are then diced into individual bases.

The main technical challenge in the fabrication of the package is the
stringent alignment tolerance of the VCSEL with respect to the fiber location.
Since the light from the VCSEL is highly collimated, the VCSEL must be
placed radially within $10~\mu$m of the center of the fiber for good
light coupling efficiency. 
The distance between the VCSEL and the end of the fiber is much less
critical and we choose 200~$\mu$m for the design shown in Fig.~\ref{fig:design}.

The alignment tolerance for the PIN diode is not stringent because the
light sensitive area of the PIN is large compared with the silica core
of the fiber.
Consequently, we have demonstrated the principle of the design by
prototyping the VCSEL optical package only.

For the placement of the VCSEL and PIN diodes, a small drop of conducting
epoxy is deposited on a trace at the location of interest.
A diode is then placed on top of the epoxy and aligned. 
Since the location of the fibers in an MT connector are defined precisely
with respect to the two guide pins, one can align the diode with respect
to the center of the two holes in the receptor under an optical comparator.
Alternatively, one can mount an MT connector with fibers attached on the
optical package and shine light into the other end of the fibers.
With the MT connector pulled back slightly from the optical package,
one can see a bright spot on the face of the diode.
The diode should then be moved so that the most sensitive area is in
the center of the beam spot.

There are many design variations depending on the requirements of an experiment.
If only one or two channels are needed in an optical package, one can
use the much smaller mini-MT connector.
A typical MT connector has either 8 or 12 fibers in a row with 250~$\mu$m
between the centers of adjacent fibers.
The number of VCSEL/PIN diodes that can be placed on a base depends on the
physical dimension of the diodes.
In the example shown in Fig.~\ref{fig:design}, one can place six diodes,
thus only half of the fibers are used.
Alternatively, one can fabricate MT connectors with wider fiber spacing
to accommodate the physical size of the diodes.
There are MT connectors that can connect up to 72 fibers in a 6 $\times$
12 matrix~\cite{tolerances}.
By staggering the diodes instead of arranging them in a row to circumvent
the diode physical size limitation, one can place more diodes in a package.
If only a small number of diodes is used in a package, one can add a wide
ground trace between two signal traces to act as a ground plane for shielding
to reduce potential crosstalk.
One can of course use a mixture of VCSEL and PIN diodes in the same package
so that the package provides two-way communication.
One can also use a common cathode VCSEL or PIN array in the package instead
of individual diodes.
In this case, the number of traces is reduced by half and a trace is added
for the common cathode connection.

\section{Prototype Results}
We fabricated several prototype VCSEL optical packages in order to test
the principle of the design.
As stated in the previous section, we only demonstrated the principle on
the VCSEL optical package and not on the PIN package, due to the more
stringent VCSEL alignment tolerances.
We used VCSELs manufactured by Truelight~\cite{Truelight}.
The physical dimension of the VCSEL was
$280~\mu {\rm m} \times 280~\mu {\rm m}$, hence we could only place six
VCSELs on a base for use with a 12-fiber MT connector as shown in
Fig.~\ref{fig:design}.
In principle, we could fabricate the U-shaped receptor using mold injection.
However, this would be somewhat time consuming and not cost effective
for a prototype test.
We therefore cut an MT connector into three pieces and machined out a pocket
for the base in each piece.
This automatically guaranteed that the guide pin hole diameter and
separation were within the mechanical specification.
In fact, this is a viable alternative to mold injection for any project
requiring only a few hundred packages.
Care must be exercised in the machining or else the plastic receptor could crack.

\begin{figure}
\begin{center}
\includegraphics*[width=10cm]{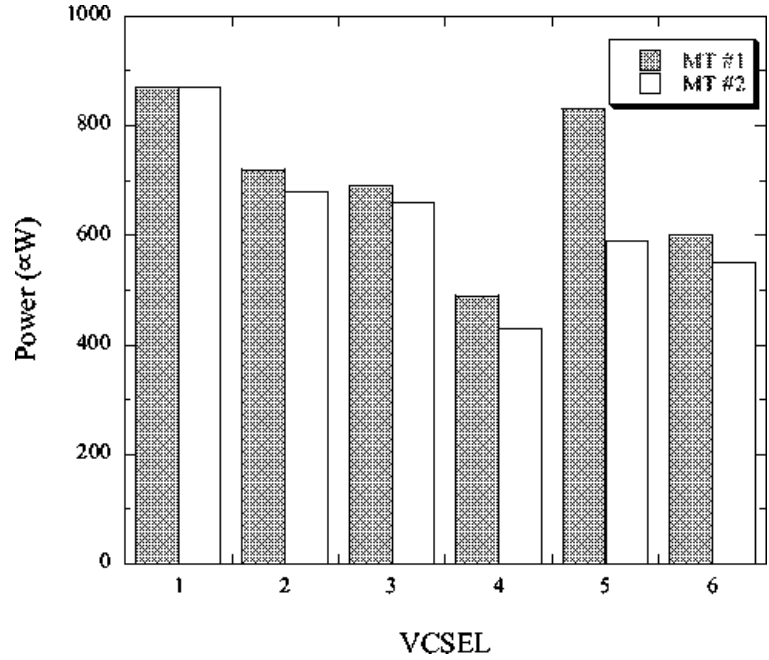}
\end{center}
\caption{
Coupled optical power of various VCSELs in two prototype optical packages.}
\label{fig:power}
\end{figure}

\begin{figure}
\begin{center}
\includegraphics*[width=10cm]{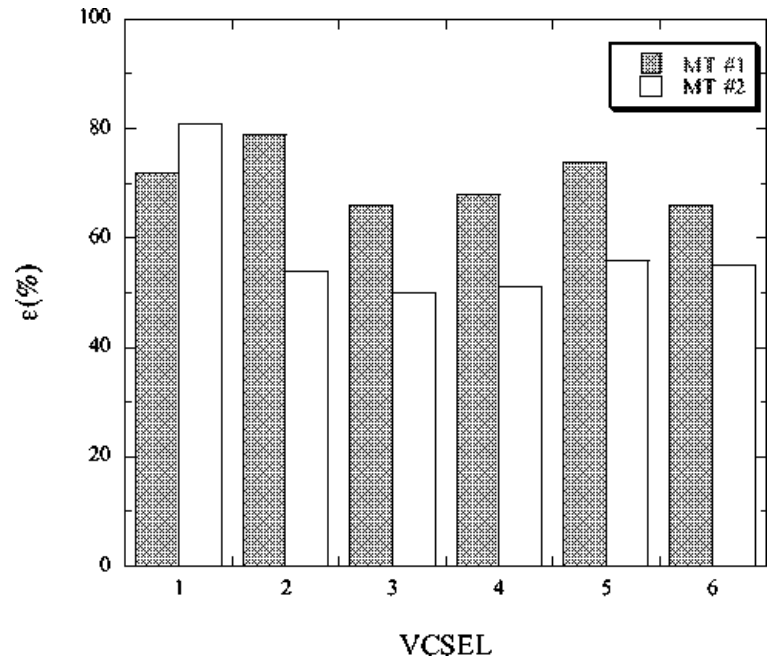}
\end{center}
\caption{
Optical power coupling efficiency of various VCSELs in two prototype optical packages.}
\label{fig:eff}
\end{figure}

\begin{figure}
\begin{center}
\includegraphics*[width=10cm]{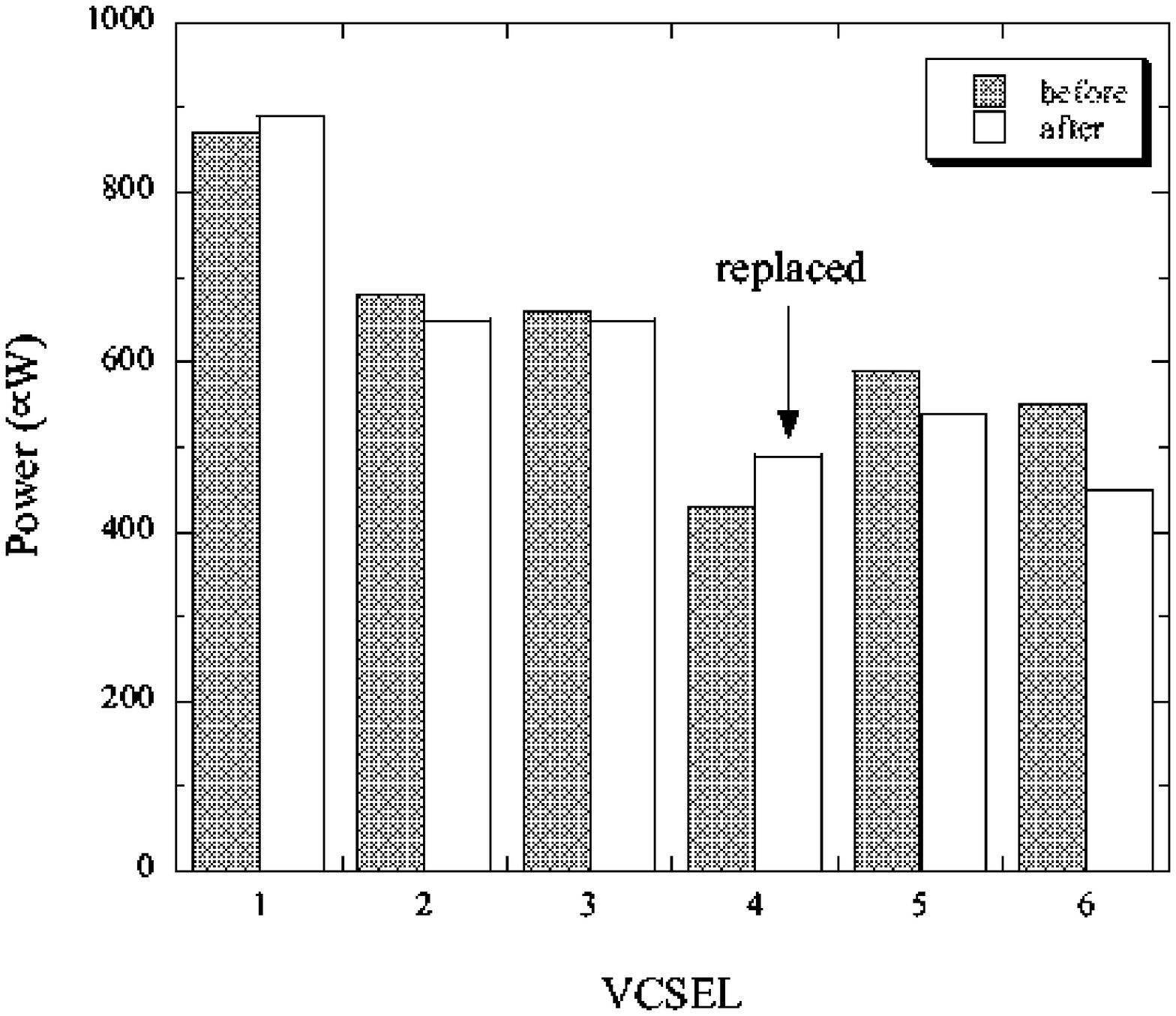}
\end{center}
\caption{
Coupled optical power of various VCSELs before and after the
replacement of the indicated VCSEL in a prototype optical package.}
\label{fig:repair}
\end{figure}

Figure~\ref{fig:power} shows the coupled VCSEL power for a 10~mA current
through the VCSEL for two prototype packages.
All channels have good coupled power, 400~$\mu$W or above.
The power coupling efficiency is shown in Fig.~\ref{fig:eff}.
All channels have good efficiency, 50\% or higher.
We have also tested the feasibility of replacing one of the VCSELs
as may be needed in a production when a VCSEL is damaged or misaligned.
Figure~\ref{fig:repair} shows the coupled optical power of all
VCSELs before and after the repair.
It is evident that the coupled power of the VCSELs after the repair are
consistent with those before the operation.
This demonstrates that a VCSEL can be replaced without damaging other channels.

\begin{figure}
\begin{center}
\includegraphics*[width=10cm]{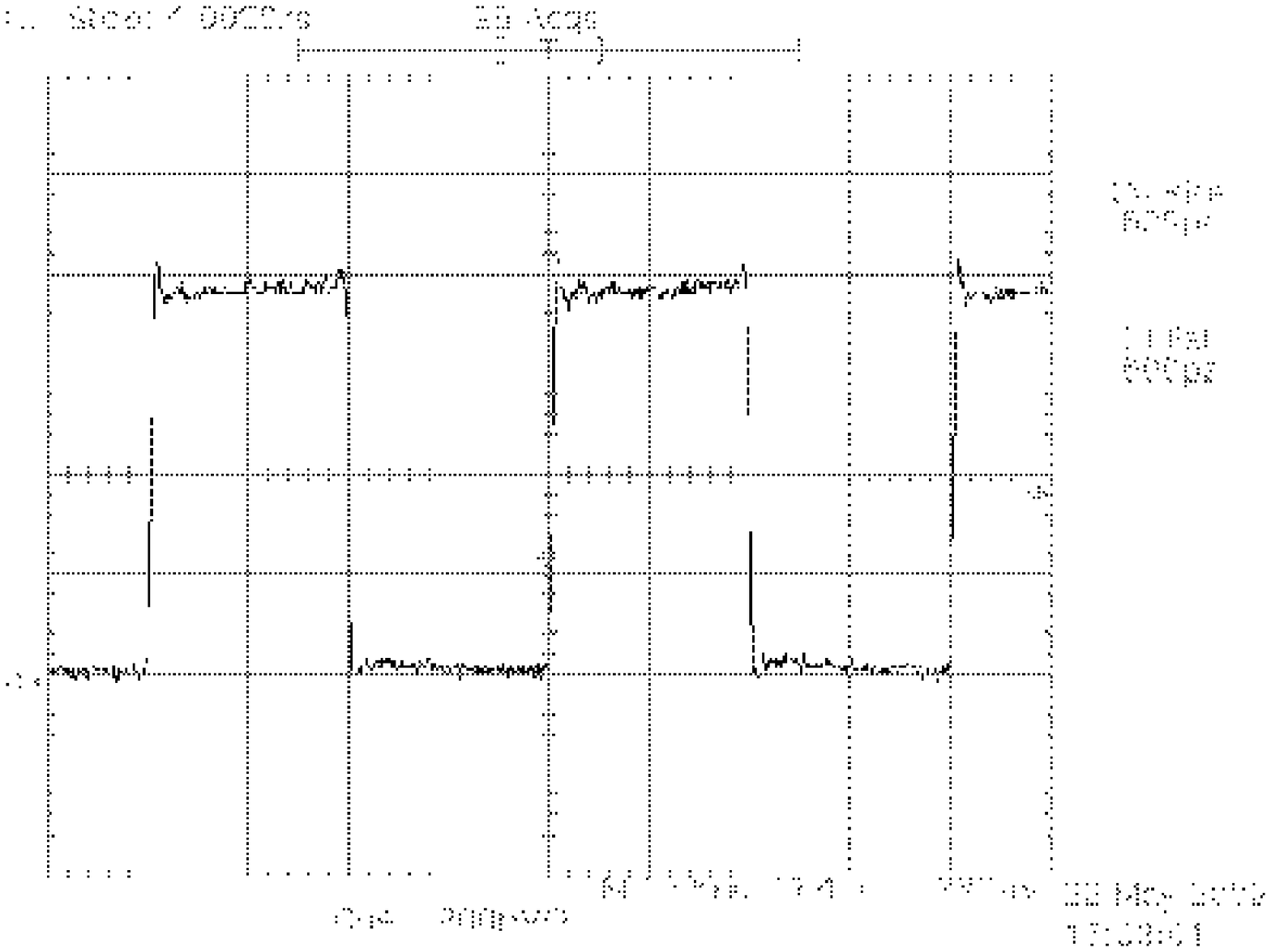}
\end{center}
\caption{
Optical signal from one of the VCSELs in a prototype optical package.}
\label{fig:waveform}
\end{figure}

The optical signal from the VCSEL is of high quality as shown in
Fig.~\ref{fig:waveform}.
Both the rise and fall times of the signal are fast, $\sim 600$~ps.

\section{Summary}
In summary, we have demonstrated the principle of an MT-style optical package.
The VCSELs in the optical package have good coupled power (efficiency).
The package is quite compact and individual VCSEL or PIN diodes in the package
can be replaced if necessary.
This package design simplifies the testing and assembly of the optical components
because the MT connector with the long fibers attached can be remounted
with ease while preserving good light coupling efficiency.

\section{Acknowledgement}
The author wishes to thank K. Arms, J. Burns, R.D.~Kass, S. Smith,
and R. Wells for their contributions.




\begin{thebibliography}{00}




\bibitem{tolerances} See for example, http://www.xanoptix.org/connectorsspecs.htm.

\bibitem{HT} Hybrid-Tek Inc., 1 Hytek Corporate Ctr, Rte. 526, Clarksburg, NJ 08510, USA.

\bibitem{Truelight} Truelight Corporation, Hsinchu, Taiwan.

\end{thebibliography}
\end{document}